\newcommand{\tri}{\triangle}
\newcommand{\rd}{{\rm d}}

\newcommand{\bI}{{\bf I}}
\newcommand{\Iav}{I_{\rm av}}

\newcommand{\Dilog}{{\rm Dilog}}

\newcommand{\be}{\begin{equation}}
\newcommand{\bea}{\begin{eqnarray}}
\newcommand{\ee}{\end{equation}}
\newcommand{\eea}{\end{eqnarray}}
\newcommand{\qa}{\alpha}
\newcommand{\qb}{\beta}
\newcommand{\qg}{\gamma}
\newcommand{\qG}{\Gamma}
\newcommand{\qd}{\delta}

\newcommand{\qe}{\varepsilon}

\newcommand{\qz}{\zeta}

\newcommand{\qh}{\eta}

\newcommand{\ql}{\lambda}

\newcommand{\qr}{\rho}
\newcommand{\qs}{\sigma}
\newcommand{\qS}{\Sigma}

\newcommand{\qf}{\varphi}

\newcommand{\qj}{\psi}

\newcommand{\Erfc}{{\rm Erfc}}

\newcommand{\tr}{{\rm tr}\;}

\newcommand{\inv}{^{-1}}

\newcommand{\intgr}[3]{\int_{#1}^{#2}\!\!{\rm d}#3\;}

\newcommand{\fr}[2]{{\textstyle \frac{#1}{#2}}}
\newcommand{\half}{\mbox{$\textstyle \frac{1}{2}$}}

\newcommand{\naar}{\rightarrow}

\newcommand{\nn}{\nonumber}

\newcommand{\prob}[1]{{\rm Prob}[#1]}

\newcommand{\Nreg}{N_{\rm reg}}
\newcommand{\vw}[1]{\left\langle #1\right\rangle}
\newcommand{\vwf}[1]{\left\langle #1\right\rangle_{\varphi}}
\newcommand{\vwe}[1]{\left\langle #1\right\rangle_{\varepsilon}}
\newcommand{\vwef}[1]{\left\langle #1\right\rangle_{\varepsilon,\varphi}}

\documentclass[11pt]{article}
\title{The entropy of keys derived from laser speckle}

\author{B. \v{S}kori\'{c}
\\Philips Research Europe
}

\date{ }

\usepackage{graphicx}
\usepackage{epsfig}
\usepackage{amssymb}
\usepackage{a4wide}

\begin{document}
\maketitle


\noindent
{\bf Abstract}\newline
{\it
Laser speckle has been proposed in a number of papers as a high-entropy source of unpredictable bits for use in security applications.
Bit strings derived from speckle can be used for a variety of security purposes such as identification,
authentication, anti-counterfeiting, secure key storage, random number generation and tamper protection.
The choice of laser speckle as a source of random keys is quite natural, 
given the chaotic properties of speckle.
However, this same chaotic behaviour also causes reproducibility problems.
Cryptographic protocols require either zero noise or very low noise in their
inputs; hence the issue of error rates is critical to applications of laser speckle in cryptography.
Most of the literature uses an error reduction method based on Gabor filtering.
Though the method is successful, it has not been thoroughly analysed.

In this paper we present a statistical analysis of Gabor-filtered speckle
patterns.
We introduce a model in which perturbations are described as random phase
changes in the source plane. Using this model we compute the second and fourth
order statistics of Gabor coefficients.
We determine the 
mutual information between perturbed and unperturbed Gabor coefficients
and the bit error rate in the derived bit string.
The mutual information provides an absolute upper bound on the number of secure bits that can be reproducibly extracted from noisy measurements. 
}

\vskip2mm
\noindent
{\bf Keywords:} Physical Unclonable Function, PUF, speckle, Gabor transform, entropy, key extraction, fuzzy extractor

\section{Introduction}

\subsection{Key generation from speckle patterns}

In \cite{PappuThesis,PappuScience} Pappu et al. proposed to use speckle patterns obtained from
coherent multiple scattering in a token to authenticate persons and
devices.
In a typical scenario,
a person carries an authentication token consisting of a transparent material with
scattering particles inside, e.g. glass with air bubbles.
When he wants to get access to some service, he presents his token to a reader
device. The device shines laser light onto the token under some predetermined
conditions (wave length, angle, focal distance, beam shape etc).
This is called a `challenge'.
The resulting speckle pattern (in transmission or reflection) under some
predetermined angle is recorded by the device. The recorded image is called the `response'.
The response is processed, e.g. by Gabor filtering, to yield a bit string that is 
reasonably insensitive to noise in the image.
This bit string is compared to a previously enrolled bit string.
If the strings are sufficiently similar, the token is
authenticated.
(A variant of this procedure, involving a scan over a length of paper, was developed
in \cite{Cowburn} for the authentication of paper documents.)

Often one cannot trust the reader device and/or the link between the reader and
the verifier. In that case a Challenge-Response Pair (CRP) cannot be used safely
more than once in the above scenario.
In order to have a secure and practical token, it must be possible to obtain many
different CRPs from one token. It must also be very hard to predict a CRP
given previously observed CRPs.
Further security requirements follow if one demands that it must be hard for an
attacker to (i)
extract all CRPs from a token in a short amount of time, and (ii) to extract
enough information from the token either to physically clone it
or to 
successfully compute its responses.
Pappu introduced the name `PUF' for a token (not necessarily optical) that satisfies all these
security requirements.
PUF stands for {\em Unclonable Physical Function}. Alternative names in the literature
are Physical One-Way Function (POWF) and Physical Random Function.
The word {\em function} stems from the fact that a response can be regarded as the evaluation of a complicated
function of the argument; the function is
parametrised by the physical structure of the token.
It turns out that the physics of multiple scattering is compatible with all PUF
requirements, especially if the token is created by a random mixing procedure of sufficiently small particles.

Going one step further than the simple matching procedure of \cite{PappuThesis,PappuScience},
it is possible to use a token's response as a secret key in a cryptographic protocol. 
This is nontrivial, since any amount of noise is fatal to ordinary cryptographic
primitives.
Secure forms of error correction, in which the redundancy data does not
leak (much) information on the secret key, were developed in \cite{JW99,LT03,DRS04}.
These techniques are called {\em fuzzy extractors} or {\em helper data schemes}.
Their application to optical PUFs was studied in \cite{STO05,HTDAI,SPTinMDM}.
Key generation from CRP measurements is an enabler for a wide variety of
security applications such as authentication, brand protection, tamper protection, anti-counterfeiting,
secure key storage and special forms of authenticated computation
\cite{GassendCPUF}.
For an overview of the subject of {\em security with noisy data} we refer to
\cite{SwNDL}.

In all of these examples, it is important to have a good understanding of the
number of random bits that can be extracted from the measurements.
Overestimation can lead to serious cryptographic weaknesses.
Underestimation leads to waste of resources.
A general framework for the computation of {\em measurement entropy} was set up in
\cite{TSSAO05} and applied to transmissive optical PUFs.

A different approach was taken in \cite{ISIT2006}, where Gabor-filtered speckle patterns
were compressed using the Context Tree Weighting (CTW) method.
The size of the compressed data gives an upper bound on the entropy.

A second important point is a good understanding of the noise that occurs in the response when the same challenge is applied multiple times.
This noise determines how much of the
total entropy of a response can be extracted in a reproducible way.
Measurement noise is caused by many factors:
temperature, moisture, stray light, mechanical misalignment, differences between
reader devices, ageing etc.
In \cite{STO05} several methods were proposed to deal with noise in
optical PUFs, e.g. alignment methods and efficient protocols.
In \cite{ISIT2006} CTW compression was employed to estimate the {\em mutual
information}
between two (Gabor-filtered) noisy measurements of the same response. This information-theoretic quantity
captures the shared entropy between two data sets and gives an upper limit
on the length of the shared key that can be reproducibly extracted from these sets.

That work has resulted in a lot of practical know-how, sufficient to set up a secure key extraction system.
What is lacking, however, is a theoretical understanding of the effects of measurement
noise on the Gabor coefficients.
In this paper we will address the issue of random perturbations and the
statistical properties of Gabor-transformed speckle patterns.

\subsection{Binarized Gabor coefficients}
\label{secGabor}

Bit strings can be extracted effectively by using a Gabor transform.
This method was proposed in \cite{PappuThesis} and further studied in \cite{STO05,ISIT2006}.
Gabor Transforms are well suited since they are insensitive to small changes in an image 
and they reveal the locations as well as the orientations of structures at different spatial frequencies.
They are used in a wide range of applications, such as 
iris recognition \cite{Daug2004},
texture analysis and image enhancement, coding and compression.

Here we briefly review the method used in the literature.
A laser beam illuminates an object and, either by transmission or reflection, produces a speckle pattern. An image of the pattern is recorded in the `detection plane'. A point in the detection plane is denoted as two-dimensional vector $\vec x$. The light intensity in the detection plane is denoted as~$I(\vec x)$.

A two-dimensional Gabor basis function $\qG(w,\vec k,\vec x_0,\vec x)$ is the product 
of a plane wave with wave vector $\vec k$ and a Gaussian with width $w$ centered on $\vec x_0$. 
We write the Gabor basis functions $\qG_{\rm IM}$ and the Gabor coefficients $G$ as follows:
\bea
\label{defG}
    G(w,\vec k,\vec x_0) &=&
    \int\! {\rd}^2 x\; \qG_{\rm IM}(w,\vec k,\vec x_0, \vec x)I(\vec x)
    \\
    \qG_{\rm IM}(w,\vec k,\vec x_0, \vec x) &=&
    \frac{1}{2\pi w^2}\sin \vec k\cdot (\vec x-\vec x_0)
    \exp[-\frac{(\vec x-\vec x_0)^2}{2w^2}].
\label{defGIM}
\eea
Only the imaginary (sine) part of the transform is considered.
This is motivated by the property that the imaginary part is invariant under spatially constant perturbations of the intensity.

Gabor coefficients $G$ are evaluated for a subset of parameters $w$, $\vec k$, $\vec x_0$, 
e.g. on a sub-lattice of positions $\vec x_0$, for two perpendicular choices of $\vec k$ 
(with equal modulus $|\vec k|$), and for one fixed~$w$.
Since the basis functions form an overcomplete set, such a restricted
choice of parameters can capture almost all information available in a speckle
image.

Coefficients are discarded if they do not exceed a certain threshold~$T$, i.e. one only keeps $|G|>T$.
The chosen coefficients are called `robust', because they are unlikely to be affected by noise.
Finally, the robust coefficients are binarized; positive values are mapped to `1' and negative to `0'.

This procedure is applied to the speckle pattern photographed during enrollment, 
and again to the image obtained in the authentication measurement.
The Hamming distance (number of bit flips) between the enrolled bitstring and 
the second bitstring depends on the threshold~$T$ and the amount of measurement noise.
Based on knowledge about the expected number of bit flips, one applies an error 
correction scheme that can cope with the noise.
It is important to keep in mind that the robust bitstring can be deceptively long. 
The actual amount of information contained in it can be much less than the length, 
due to correlations between the Gabor coefficients \cite{STO05}.


\subsection{Contributions and outline of this paper}

In Section~\ref{secrandomphase} we first briefly 
describe the random phase model
and the intensity statistics that are obtained from it.
We motivate our use of this model.
In Sections \ref{secGstat} and~\ref{secpert} we study the statistics of Gabor
coefficients and the effects of random perturbations.
This paper contains the following novel contributions:
\begin{itemize}
\item
In Section~\ref{sechighermoments}
we analyse the statistical properties of the Gabor-transform (\ref{defG}) of a speckle pattern. 
We present a procedure for computing arbitrary moments of the distribution.
Computation of the first four moments shows that there are small deviations from
the normal form.
\item
In Section~\ref{secdetectornoise} we compute the information content of a set of Gabor coefficients, for a given noise level of the detector. 
The entropy per typical speckle area turns out to be proportional to the {\em square} of the logarithm of the signal to noise ratio.
\item
In Section~\ref{secdefpert}
we introduce a method of perturbing a speckle pattern in the random phase model. 
Each $\ql^2$ sized source region has its phase shifted by a small random amount~$\qe$, where $\qe$ is drawn 
from a uniform distribution of width~$2q$. 
By tuning $q\in[\half\tri\qf,\pi]$ (where $\tri\qf$ denotes the phase uncertainty due to the 
number-phase uncertainty relation), 
the magnitude of the perturbation is selected. 
We use this kind of perturbation to represent a misalignment, such as a shift or rotation of the token or the laser, 
or a change in the structure of a token.
\item
In Section~\ref{secmutual} we compute the mutual information between the original speckle
`source' and the perturbed one as a function of the noise strength~$q$. 
The result is proportional to $\log (\pi/q)$.
\item
In Section \ref{secperturbI} we show that the correlation between the intensity
before and after a perturbation is given by $\frac{\sin^2 q}{q^2}$.
In Section \ref{secpertgabor} the same correlation is obtained between Gabor
coefficients before and after a perturbation.
\item
In Section~\ref{secIGpert} we compute the mutual information between a set of Gabor coefficients before and after a perturbation.
\item
In Section~\ref{secbitflip} we compute the bit flip probability
for a binarized Gabor coefficient due to a random perturbation.
\item
In Section~\ref{secexperiments} we give experimental results.
(i) It turns out that the empirical distribution function of the Gabor coefficients is consistent with theory.
There is a noticeable deviation from the Gaussian form.
The theoretical prediction of the variance matches very well with the data.
(ii) We studied random perturbations by doing measurements on a sample whose surface structure slowly changes in time. Correlations were determined between the state before and after a perturbation.
As expected from the theoretical results, there is a linear relation between the correlation function of the intensity and the correlation function of the Gabor coefficients.

\end{itemize}


\section{The random phase model}
\label{secrandomphase}
\subsection{Motivation and definitions}
\label{secmotivation}

Throughout this paper we use the random phase model as described by Goodman~\cite{Goodman}, 
in the Fresnel approximation, in a free space geometry, for completely polarised light.
This model has the advantage of being relatively simple while yielding intensity statistics 
that agree with experimental observations.
We depart from the traditional approach only in one respect. In~\cite{Goodman} 
the components of the electric field amplitude ($A_x$, $A_y$) in the detection plane are 
sometimes treated as the `fundamental' degrees of freedom.  
For instance, second order and higher order intensity correlations 
are
derived using the Gaussian 
distribution of $\vec A(\vec x)$.
The traditional approach has the drawback that it is very difficult to keep track of the number of degrees of freedom: 
It looks as if there is one degree of freedom per (continuum!) location $\vec x$ in the source plane.
In fact the physical degrees of freedom lie in the 
source plane (defined as the exit plane of a transmissive PUF, or as the surface that reflects the laser light) and they are very easy to identify and to count.
Another drawback is that there is no natural way to introduce misalignment perturbations 
in terms of the $\vec A(\vec x)$ variables. 
On a more esthetic level, there is the drawback of having to rely on the Central
Limit Theorem to get the Gaussian distribution of $\vec A$, while in fact the number of 
random amplitudes added together is not infinite but merely very large.

For these reasons we base our calculations on the random phases in the source plane as the fundamental degrees of freedom. 
All the well known speckle properties are of course reproduced in this approach.
The model looks as follows.
Diffused light leaves the PUF at the exit plane, through a disc-shaped region with radius $R$ which we call the `source'. 
We assume that the intensity is the same everywhere in the source. (We normalize the intensity to~1).
Hence the source is modelled as a collection of random phases~$\qf$.
The disc is divided into small regions of area~$\ql^2$.
(The total number of regions is denoted as $\Nreg=\pi R^2/\ql^2$).
Together these generate the speckle pattern according to Huygens' principle.
The complex amplitudes $\qa$ in each region are the basic degrees of freedom. We write
\be
	\qa_{\vec a} = \exp i\qf_{\vec a},
\ee
where the subscript $\vec a$ denotes a discrete two-dimensional coordinate in the source, with $|\vec a|<R$. 
The phases $\qf_{\vec a}$ at all the locations $\vec a$ are independent stochastic variables, with a 
uniform distribution in the interval $(-\pi,\pi]$.
We introduce the notation $\vwf{\cdot}$ for taking the expectation value with respect to the random phases.
We have
\bea
	\vwf{(\qa_{\vec a})^n}=0 \quad{\rm for}\; n\in{\mathbb N}^+
	&;&
	\vwf{\qa_{\vec a} \qa_{\vec b}^*}=\qd_{\vec a,\vec b}.
\label{vwqaqa}
\eea
From these basic rules it is straightforward to derive many-point correlations.

Note that in adopting {\em independent} random phases we ignore the correlations that are known to exist between the phases as a consequence of either
(a) multiple coherent scattering in a diffusive medium (see e.g. \cite{FengPRL}),
or
(b) height and/or orientation correlations between microscopic pieces of a rough surface. 
These correlations lead to a reduction of the number of degrees of freedom. However, in this paper we are primarily interested in the influence that the parameter choices in (\ref{defGIM}) have on the information that can be extracted from a Gabor-filtered speckle pattern.
In this context the correlations between phases in the source plane are only of minor importance. Hence we will ignore them and work
with~(\ref{vwqaqa}).

The distance between the source and the detection plane is denoted as~$z$.
We assume $z\gg\ql$ and use the Fresnel approximation. For the complex amplitude $A=A_x+iA_y$ in a point 
$\vec x$ in the detection plane we then have
\be
	A(\vec x)=\frac{\ql}{z}\sum_{\vec a}\qa_{\vec a}
	\exp -i\frac{\pi}{\ql z}(\vec x-\vec a)^2.
\label{amplitude}
\ee


\subsection{Entropy of the source}
\label{secentropy}

The entropy of the source is an upper bound on the entropy of the speckle pattern in a half-
sphere.
The entropy of the source is easily computed in the random phase model.
The phase distribution is completely uniform. Therefore the entropy reduces to the 
logarithm of the number of possible states. 
It is well known that a coherent state with photon number $N_0$ has an uncertainty 
$\tri N=\sqrt{N_0}$ in the photon number.
Using the number-phase uncertainty relation $\tri N\tri\qf=\half$, we find a phase 
discretisation $\tri\qf = 1/(2\sqrt{N_0})$.
Hence the entropy (expressed in bits) is given by
\be
	H[\qa] = \log_2 (\frac{2\pi}{\tri\qf})^{\Nreg}=
	\Nreg\log_2 (4\pi\sqrt{N_0}).
\label{entropy}
\ee
Note that this result is equivalent to the estimate in~\cite{STO05}, where the light 
exiting the PUF was described in terms of transversal momentum modes.
In~\cite{STO05} the correlations between modes were studied as well, and the 
resulting entropy reduction was estimated. In this paper we will not take such correlations into account.

In order to get some feeling for the orders of magnitude we substitute numbers into~(\ref{entropy}).
A laser with $\ql=780$nm
produces an output power $P=$1mW, and a measurement with a CCD camera 
takes about $\tri t=$1ms. The total number of photons involved in one measurement is 
$P\tri t/(hc/\ql)$, where $h$ is Planck's constant and $c$ is the velocity of light.
Thus we arrive at $N_0=\ql P\tri t/(hc\Nreg)$.
Assuming a source diameter of 1mm, we get $\Nreg=1.3\cdot 10^6$ and $N_0=3\cdot 10^6$ 
photons per region, yielding an entropy of approximately 14 bits per region of size $\ql^2$.


\subsection{Statistics of the intensity}
All the well known statistical properties of the intensity can be derived from the random phase model.
For completeness and for use in later sections, we briefly discuss how these properties are derived.
The intensity at position $\vec x$ in the plane of detection is given by the squared modulus 
of the amplitude (\ref{amplitude}),
\be
	I(\vec x)=\left|A(\vec x)\right|^2 =
	\frac{\ql^2}{z^2}\sum_{\vec a,\vec b}
	\qa_{\vec a}\qa^*_{\vec b} \; \exp\frac{i\pi}{\ql z}
	\left[\vec b^2-\vec a^2+2\vec x\cdot(\vec a-\vec b)\right].
\label{intensity}
\ee
The average intensity $\Iav$ is obtained by taking the expectation value $\vwf{\cdot}$ and 
directly applying~(\ref{vwqaqa}).
Summation $\sum_{\vec a}$ over a constant yields a factor $\Nreg$. We obtain
\be
	\Iav=\vwf{I(\vec x)} = \frac{\Nreg\ql^2}{z^2} 
	=\frac{\pi R^2}{z^2}. 
\ee
For second order statistics of the intensity one needs 4th order correlations of the random 
amplitudes~$\qa$. In particular, from (\ref{vwqaqa}) it follows that
\be
	\vwf{\qa_{\vec a_1}\qa_{\vec a_2}\qa^*_{\vec b_1}\qa^*_{\vec b_2}}
	= \qd_{\vec a_1,\vec b_1}\qd_{\vec a_2,\vec b_2}
	+\qd_{\vec a_1,\vec b_2}\qd_{\vec a_2,\vec b_1}.
\label{vwqa4}
\ee
Using (\ref{vwqa4}), the well known results follow for the
variance ($\qs_I=\Iav$) and for the intensity correlation function $C_I$,
\be
	C_I(\vec x,\vec x')  :=  
	\frac{\vwf{I(\vec x)I(\vec x')}-\Iav^2}
	{\qs_I^2}=
	4\left[ \frac{J_1(|\vec x'-\vec x|/M)}
	{|\vec x'-\vec x|/M} \right]^2,
\label{CIxxp}
\ee
where $M$ is a constant proportional to the average speckle size,
\be
	M=\frac{\ql z}{2\pi R}
\label{defM}
\ee
and $J_1$ is a Bessel function.
Higher order expectation values can also be computed. In particular, in order to derive 
the well known exponential probability density 
\be
	p(I)=\Iav\inv\exp(-I/\Iav), 
\label{pI}
\ee
it has to be shown that
$\vwf{[I(\vec x)]^n}=\Iav^n n!$. This is done using the following correlation function,
which also follows from (\ref{vwqaqa}),
\be
	\vwf{\qa_{\vec a_1}\ldots\qa_{\vec a_n}\cdot
	\qa^*_{\vec b_1}\ldots\qa^*_{\vec b_n}}=
	\sum_{\nu\in S_n} \prod_{k=1}^n 
	\qd_{\vec a_k,\vec b_{\nu(k)}}.
\label{vwqanqan}
\ee
Here $S_n$ stands for the `symmetric group' of all the $n!$ possible permutations of the numbers $1\ldots n$. 
Eq.~(\ref{vwqanqan}) is also used in the derivation of the joint probability 
distribution $p(I(\vec x),I(\vec x'))$. This distribution follows from the expectation value
\be
	\vwf{[I(\vec x)]^n [I(\vec x')]^m} =
	\Iav^{n+m} n! m! \;\;{}_2F_1(-n,-m; 1; C_I(\vec x,\vec x'))
\label{momentInm}
\ee
and is given by
\be
	p(I(\vec x),I(\vec x'))=\frac{1}{\Iav^2 (1-C_I)}
	\exp\left[ -\frac{I(\vec x)+I(\vec x')}{\Iav(1-C_I)} \right]
	I_0\left(\frac{2\sqrt{I(\vec x)I(\vec x')}}{\Iav}
	\frac{\sqrt{C_I}}{1-C_I}
	\right).
\label{pIIp}
\ee
Here $C_I$ is shorthand notation for $C_I(\vec x,\vec x')$ and $I_0$ is a Bessel function.


\section{Statistics of the Gabor coefficients}
\label{secGstat}

\subsection{Second order statistics}
\label{secGstat2ndorder}
Second order statistics of the Gabor coefficients of a speckle pattern
were calculated in~\cite{STO05}.
In this section we study the higher order statistics.
In particular we show that all the odd moments are zero and that the fourth moment
is dominated by the Gaussian contribution; i.e. the probability distribution
of a Gabor coefficient is `almost' Gaussian.

We will often use shorthand notation $G$ for $G(w,\vec k,\vec x)$.
From the fact that $\qG_{\rm IM}$ (\ref{defGIM}) is an odd function in $(\vec x-\vec x_0)$ 
it is easily seen that $\vwf{G}=0$.
In~\cite{STO05} it was shown that the variance of $G$ is given by
\be
	\qs^2_G(w,\vec k,\vec x_0)\approx
	\Iav^2 (1-2\qg) e^{-(1-\qg)w^2 k^2}
	\sinh \qg w^2 k^2,
\label{sigG}
\ee
where $\qg=\half [1+M^2\qS^2/(2w^2)]\inv$
and
$\qS\approx 1.29$ is a numerical constant.
The constant $\qS$ originates from an approximation of the correlation (\ref{CIxxp})
by a Gaussian curve.
The correlation between two different Gabor coefficients
was also computed.
The correlation is defined as
\be
	C_G(w,w',\vec k,\vec k',\vec x,\vec x')
	:=\frac{\vwf{G(w,\vec k,\vec x)G(w',\vec k',\vec x')}}
	{\qs_G(w,\vec k,\vec x)\qs_G(w',\vec k',\vec x')},
\label{defCG}
\ee
and the following result was obtained for $w'=w$,
\bea
	C_G&\approx&\exp\left[-\frac{\qg}{2}\cdot \frac{(\vec x'-\vec x)^2}
	{w^2}\right]\times
	\nn\\ &&
	\frac{e^{\qg w^2\vec k\cdot\vec k'}
	\cos[\qg(\vec x'-\vec x)\cdot(\vec k'+\vec k)]
	-e^{-\qg w^2\vec k\cdot\vec k'}
	\cos[\qg(\vec x'-\vec x)\cdot(\vec k'-\vec k)]}
	{2\sqrt{\sinh\qg w^2 k^2}\sqrt{\sinh\qg w^2 k'^2}}.
\label{CG}
\eea
The result (\ref{CG}) is accurate for small distances $|\vec x'-\vec x|$.
For larger distances, the Gaussian tail underestimates the actual correlation.

\subsection{Computation of higher moments}
\label{sechighermoments}

We present a procedure that allows for the computation of arbitrary moments of~$G$.
Substitution of (\ref{intensity}) into (\ref{defGIM}) gives an expression for $G$ in terms of the random phases,
\bea
	G(w,\vec k,\vec x_0) &=&
	\frac{i\ql^2}{z^2}e^{-\half w^2 k^2}\sum_{\vec a,\vec b}
	\qa_{\vec a}\qa^*_{\vec b}
	\exp \frac{i\pi}{\ql z}[\vec b^2-\vec a^2
	+2\vec x_0\cdot(\vec a-\vec b)]
	\nn\\ &&
	\exp \left[ -\frac{(2\pi)^2}{(\ql z)^2}(\vec a-\vec b)^2\right]
	\;\sinh\left[\frac{2\pi}{\ql z}w^2 \vec k\cdot(\vec a-\vec b)\right].
\label{GIMqa}
\eea
Taking the $n$th power of (\ref{GIMqa}) and averaging over the random phases leads to 
a correlation function of the form (\ref{vwqanqan}).
Each $\vec b_j$ evaluates to $\vec a_{\nu(j)}$.
For each permutation, the sums $\sum_j(\vec a_j-\vec b_j)$ and
$\sum_j(\vec b_j^2-\vec a_j^2)$ reduce to zero.
In this way we obtain the following expression
\bea
\label{Gn}
	\vwf{(G)^n} &=& \left[\frac{i\ql^2}{z^2}\right]^n
	e^{-nw^2 k^2/2}
	\sum_{\vec a_1 \cdots \vec a_n}
	\exp\left[-\frac{w^2}{M^2 R^2}\sum_{j=1}^n \vec a_j^2
	\right] 
	\\ &&
	\sum_{\nu\in S_n}
	\exp\left[\frac{w^2}{M^2 R^2}\sum_{j=1}^n
	\vec a_j \cdot\vec a_{\nu(j)}\right]
	\prod_{t=1}^n \sinh \frac{w^2}{MR} \vec k\cdot
	(\vec a_t -\vec a_{\nu(t)}).
	\nn
\eea
It is immediately clear that odd moments vanish, since (\ref{Gn}) is real-valued only for even~$n$.

We can simplify (\ref{Gn}) by noting that the product of sinh's is zero for permutations that have a fixed point.
In other words, all cycles of a permutation have to be longer than~1.
Also, we note that the $\vec a$-summation factors into a product of independent 
sums in accordance with the cycle structure of the permutation. For instance, 
the permutation $(231)(564)$ leads to the factorisation
$(\sum_{\vec a_1\vec a_2\vec a_3}\cdots)
(\sum_{\vec a_4\vec a_5\vec a_6}\cdots)$, where the dots indicate an expression 
depending only on the three denoted variables.
Furthermore, the outcome of each such factor depends only on the length of the 
cycle, and not on the identity of the summation variables.

The 4th moment is obtained as follows.
Among the $4!$ permutations of $\{1,2,3,4\}$ there are 3 containing two cycles 
of length two and 6 containing one cycle of length four.
The 3 permutations with two cycles give rise to a contribution
$3(\qs_G^2)^2$, which precisely corresponds to the 
4th moment of a Gaussian distribution. The non-Gaussian contribution from the 6 remaining permutations 
(i.e. the 4th cumulant) is computed in appendices \ref{appG4} and~\ref{appG4_2}.
In the regime $w\ll M$ it turns out that the ratio of the non-Gaussian part to the Gaussian is 1/64.
In the regime $w\gtrapprox 3M$ this ratio is of order ${\cal O}(M^2/w^2)$.

\subsection{Estimated entropy of a set of Gabor coefficients for given detector noise}
\label{secdetectornoise}

We use a Gaussian approximation for the distribution function of the Gabor coefficients
in order to derive an upper bound on the entropy of a set of Gabor coefficients, at a given
noise level of the detector.
This is a useful exercise for two reasons. First, it is not a priori clear how much of the
information present in the source ends up in the Gabor coefficients.
Second, detector noise affects the Gabor coefficients in a nontrivial way.

As mentioned in Section~\ref{secGabor}, a relatively small set of coefficients can
capture almost all the information available in a speckle image.
We will consider the example given in Section~\ref{secGabor}, which is also the
choice made in \cite{PappuThesis,STO05,ISIT2006}.
Taking more than one width $w$, or more than one wave number $|\vec k|$ does not
make much sense, since all the features in a speckle pattern have more or less the
same length scale, namely the average speckle size.
It is also not very useful to take more than two angles of the wave vector: Eq.~(\ref{CG}) shows us that there is a strong correlation 
$\sinh [\qg w^2 k^2 \cos\qz]/\sinh \qg w^2 k^2$
between Gabor coefficients at the
same position $\vec x$, when their $\vec k$-vectors have a mutual angle~$\qz$.
Hence, for small $\qz$ there is a lot of redundancy. 

We take a single width $w$, a single wave vector length $k$, two perpendicular directions
$\qj_1$, $\qj_2$, and a lattice of positions~$\vec x$.  
We introduce the following definitions:
\bea
	\qS_G^{[jj]}(\vec x,\vec x') &:=& \vwf{G(w,\vec k_j,\vec x)G(w,\vec k_j,\vec x')}
	\nn\\ & =&
	\Iav^2(\half-\qg)
	\exp\left[-\frac{\qg}{2}\frac{(\vec x'-\vec x)^2}{w^2}\right]
	\left\{e^{(2\qg-1) w^2k^2}\cos 2\qg\vec k_j\cdot(\vec x'-\vec x)-e^{-w^2k^2}\right\}
	\nn\\
	\qS_G^{[12]}(\vec x,\vec x') &:=& \vwf{G(w,\vec k_1,\vec x)G(w,\vec k_2,\vec x')}
	\nn\\ &=&
	-\Iav^2(1-2\qg)e^{(\qg-1)w^2k^2}
	\exp\left[-\frac{\qg}{2}\frac{(\vec x'-\vec x)^2}{w^2}\right]
	\nn\\ && \times 
	\sin[\qg\vec k_1\cdot(\vec x'-\vec x)]\sin[\qg\vec k_2\cdot(\vec x'-\vec x)].
\label{qSGparts}
\eea
We define the combined covariance matrix $\qS_G$ as
\be
	\qS_G=\left(\matrix{\qS_G^{[11]} & \qS_G^{[12]}\cr 
	\qS_G^{[12]} & \qS_G^{[22]}}\right).
\label{qSG}
\ee
We consider the joint probability distribution for all the Gabor coefficients to be Gaussian.
This, of course, is not true, as we see from the nonzero even moments in Section~\ref{sechighermoments}.
However, for a given mean and covariance matrix, the Gaussian distribution has a higher entropy than any other distribution.
Hence our procedure yields an upper bound on the entropy.

We furthermore assume that the detector noise is Gaussian, independent of the
intensity and independent for each pixel.
We can then apply the well known channel capacity formula (see e.g. \cite{CT91}), which
expresses the mutual information between a source and a noisy detection as the
logarithm
of a signal to noise ratio.
Let $\vec G$ be a vector of Gabor coefficients, and $\qd\vec G$ the detector noise in these 
coefficients, then
\be
	\bI(\vec G; \vec G+\qd \vec G) = \half \log_2|\det({\bf 1}+\qS_N^{-1}\qS_G)|.
\label{SNR}
\ee
Here the matrix $\qS_N$ is the covariance matrix of the noise.
We determine $\qS_N$ as follows.
The independence of the noise in each pixel yields 
\be
	\langle \qd I(\vec x)\; \qd I(\vec x')\rangle_{\rm n}=N_I^2 t\;\qd(\vec x-\vec x'), 
\label{dIdI}
\ee
where $N_I$ denotes the 
noise amplitude, $t$ is the area of a detector pixel
and the notation $\langle\cdot\rangle_{\rm n}$ denotes a noise average.
Note that $N_I$ is lower bounded by the shot noise.

Using (\ref{dIdI}) and the definition of the Gabor coefficients (\ref{defG}) we 
obtain the following expression for the covariance of the noise in the Gabor coefficients
\be
	\left\langle\qd G(w,\vec k_1,\vec x_1)\;
	\qd G(w,\vec k_2,\vec x_2) \right\rangle_{\rm n}
	= N_I^2 t\int\!\rd^2 x\; \qG_{\rm IM}(w,\vec k_1,\vec x_1,\vec x)
	\qG_{\rm IM}(w,\vec k_2,\vec x_2,\vec x).
\ee
Evaluation of the integral is straightforward (it is equivalent to the derivation
of $\qS_G$ in \cite{STO05} in the limit $\qS\naar 0$) and yields, in the special
case $|\vec k_1|=|\vec k_2|=k$,
\bea
\label{resultqSN}
	\qS_N^{[jj]}(\vec x,\vec x')&=& \hphantom{-}
	\frac{N_I^2 t}{8\pi w^2}e^{-\frac{(\vec x'-\vec x)^2}
	{4w^2}}\left[ \cos\vec k_j\cdot(\vec x'-\vec x)-e^{-w^2 k^2} \right]
	\\
	\qS_N^{[12]}(\vec x,\vec x')&=&-\frac{N_I^2 t}{8\pi w^2} 
	e^{-\frac{(\vec x'-\vec x)^2}{4w^2}} e^{-\half w^2 k^2}
	\sin[\half\vec k_1\cdot(\vec x'-\vec x)]
	\sin[\half\vec k_2\cdot(\vec x'-\vec x)].
\eea
$\qS_N$ has the same block structure as $\qS_G$.
Notice that both $\qS_G$ and $\qS_N$ depend on $\vec x$ and $\vec x'$ only through
the difference $(\vec x'-\vec x)$. This allows 
for efficient computation of the determinant in (\ref{SNR})
by diagonalisation in the Fourier domain.
Notice further that the off-diagonal blocks ($\vec k_1\perp\vec k_2$)
have very small values compared to the diagonal blocks due to the presence of
the sine factors.

We compute the Fourier transforms  as follows.
Strictly speaking, the transform is a summation over the finite $\vec x$-grid.
We denote the size of the grid as~$L$.
However, we will sum to infinity, since the error introduced in this way is only
an edge effect. The finiteness of $L$ is still reflected by the discretisation
of the momentum $\vec p$ conjugate to $\vec x'-\vec x$.
All momenta are multiples of $\pi/L$. 
The highest momentum is determined by the lattice constant $\ell$ of the $\vec x$-
grid.
As a second approximation, we will replace the summations over $\vec x$ and $\vec x'$
by integrations, i.e. $\sum_{\vec x}\naar\ell^{-2}\int\!\rd^2 x$. 
This is a good approximation provided that the lattice constant
is significantly smaller than the average speckle size.
Using this procedure we obtain
\bea
	\tilde\qS_G^{[jj]}(\vec p)&:=&
	\ell^{-2}\int\!\rd^2(\vec x'-\vec x)\;\qS_G^{[jj]}
	e^{-i\vec p\cdot(\vec x'-\vec x)}
	\nn\\ &=&
	4\pi\Iav^2 \frac{w^2}{\ell^2} 
	(\frac{1}{2\qg}-1)e^{-w^2 k^2}e^{-\frac{w^2}{2\qg}p^2}
	\sinh^2 w^2\vec k_j\cdot\vec p
\label{qSGfourier}
	\\
	\tilde\qS_N^{[jj]}(\vec p)&:=&
	\ell^{-2}\int\!\rd^2(\vec x'-\vec x)\;\qS_N^{[jj]}e^{-i\vec p\cdot(\vec x'-\vec x)}
	\nn\\ &=&
	N_I^2 \frac{t}{\ell^2}\; 
	e^{-w^2 k^2}e^{-w^2 p^2}\sinh^2 w^2\vec k_j\cdot\vec p. 
\label{qSNfourier}
\eea
Substitution of (\ref{qSGfourier}) and (\ref{qSNfourier}) into (\ref{SNR}) gives
us the mutual information for one Gabor direction,
\bea
	\bI(G^{[j]};G^{[j]}+\qd G^{[j]})&=& 
	\half\log_2\left|\det\left( 1+(\qS_N^{[jj]})^{-1}\qS_G^{[jj]} \right)\right|
	=\half\tr\log_2\left| 1+(\qS_N^{[jj]})^{-1}\qS_G^{[jj]} \right|
	\nn\\ &=&
	\half\sum_{\vec p}\log_2\left(1+c_1 e^{-c_2 p^2}\right),
\label{sumlogc1c2}
\eea
where we have defined the constants $c_1$, $c_2$ as
\bea
	c_1= (\frac{1}{2\qg}-1)\frac{4\pi\Iav^2 w^2}{N_I^2 t}
	=2\pi\qS^2\frac{\Iav^2}{N_I^2}\frac{M^2}{t}
	&\quad ; \quad&
	c_2= (\frac{1}{2\qg}-1) w^2
	=\half M^2 \qS^2.
\label{defc1c2}
\eea
Note that $c_2$ is proportional to the average speckle area,
while $c_1$ plays the role of a signal to noise ratio (SNR).
Note too that all dependence on $w$ and $\vec k$ has disappeared from~(\ref{sumlogc1c2}). 
The reason is that we are computing a generic upper bound. Computation of the actual amount of extracted information will in general depend on $w$ and~$\vec k$.

The summation domain of $\sum_{\vec p}$ is given by $(p_x,p_y)=(i,j)\frac{\pi}{L}$,
with $i,j\in{\mathbb Z}$ and $|p_x|<\pi/\ell$, $|p_y|<\pi/\ell$. 
Eq.~(\ref{sumlogc1c2}) can be further evaluated by approximating the momentum sum
by an integration. Notice that the summand in (\ref{sumlogc1c2}) only depends on
the length of $\vec p$. Hence the computation is simplified in polar coordinates,
\be
	\sum_{\vec p} f(p^2)\approx 
	(\frac{L}{\pi})^2\int_{\pi/L}^{\pi/\ell}\!\rd p_x
	\int_{\pi/L}^{\pi/\ell}\!\rd p_y\; f(p^2)
	=
	\frac{L^2}{\pi}\int_{(\pi/L)^2}^{(\pi/\ell)^2}\rd p^2 \; f(p^2).
\label{psumapprox}
\ee
Applying the approximation (\ref{psumapprox}) to (\ref{sumlogc1c2}) we obtain
\be
	\bI(G^{[j]};G^{[j]}+\qd G^{[j]})\approx
	\frac{L^2}{2\pi}\left[
	-\frac{1}{c_2}\Dilog(-\frac{\exp c_2 p^2}{c_1})-\half c_2 p^4
	+p^2\ln c_1\right]_{p=\pi/L}^{\pi/\ell}.
\label{dilog}
\ee
Here Dilog is the dilogarithm function. 
Note that (\ref{dilog}) is not expressed in bits but in natural units (`nats'), i.e. using the natural logarithm $\ln$ instead of the base-2 $\log_2$.

Eq.~(\ref{dilog}) can be further evaluated if $c_1$ is very large (we call this the `large SNR' case), or when $c_1$ is very small (`small SNR').
We define a quantity $y$ as
\be
	y=c_1 e^{-c_2\pi^2/\ell^2}.
\label{defy}
\ee
The crossover between the two regimes lies around $y=1$.

\subsubsection*{Large SNR}
In this case we have $y\gg 1$.
Applying the asymptotic relation
\be
	\Dilog(x)=x+{\cal O}(x^2)
	\quad\mbox{for}\quad |x|\ll 1
\label{dilogsmall}
\ee
to (\ref{dilog}) we obtain
\be
	\bI(G^{[j]};G^{[j]}+\qd G^{[j]})\approx
	\frac{\pi L^2}{2\ell^2}\left[
	\ln c_1 - \frac{c_2\pi^2}{2\ell^2}+{\cal O}(y^{-1})
	+{\cal O}(\frac{\ell^2 M^2}{L^4})
	+{\cal O}(\frac{\ell^2}{c_1 L^2})
	\right].
\label{IlargeSNR}
\ee

\subsubsection*{Small SNR}
In this case we have $y\ll 1$.
Using (\ref{dilogsmall}) and the asymptotic behaviour
\be
	\Dilog(-x)=-\pi^2/6-\half(\ln x)^2+{\cal O}(x^{-1})
	\quad\mbox{for}\quad x\naar\infty
\label{diloglarge}
\ee
we get
\be
	\bI(G^{[jj]};G^{[jj]}+\qd G^{[jj]})\approx
	\frac{L^2}{4\pi c_2}\left[(\ln c_1)^2+\frac{\pi^2}{3}
	+{\cal O}(y)+{\cal O}(\frac{1}{c_1})
	+{\cal O}(\frac{M^2}{L^2})
	\right].
\ee
\begin{figure}[h!]
\begin{center}
\includegraphics[width=80mm]{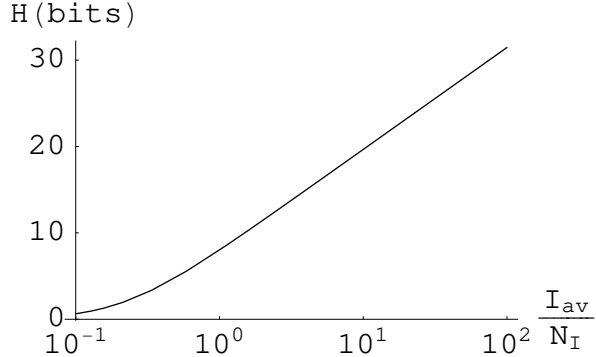}
\caption{\it Mutual information, per average speckle area, between a noiseless and a noisy vector $G^{[j]}$ of Gabor coefficients, plotted as a function of $\Iav/N_I$. The other parameters are: $L=800$ pixels, $M=3$ pixels, $\ell=5$ pixels, $t=1$ pixel\,$^2$. The radius of an average speckle has been set, somewhat arbitrarily, to~$M$.}
\label{figmutualG}
\end{center}
\end{figure}

Eq.~(\ref{dilog}) is plotted in Fig.~\ref{figmutualG} on a logarithmic scale; The factor $L^2$ was replaced by $\pi M^2$ to obtain a result per average speckle area.
We see the transition from parabolic behaviour to linear behaviour around $\Iav/N_I=1$. The linear part of the curve is the usual `log SNR' dependence, but here it occurs with a nonzero offset.


\section{Perturbing speckle patterns}
\label{secpert}

\subsection{Phase perturbations}
\label{secdefpert}

We introduce a method for perturbing a speckle pattern in the model
of Section~\ref{secmotivation}.
We shift all the phases by random amounts, 
\be
	\qf_{\vec a}\naar \hat\qf_{\vec a}=
	\qf_{\vec a}+\qe_{\vec a}.
\label{defperturb}
\ee
This is done independently for all locations~$\vec a$.
The perturbations $\qe$ are chosen from a distribution that is uniform on the interval 
$(-q,q]$ and zero elsewhere. The parameter $q$ is the `strength' of the perturbation. 
The maximum value is $q=\pi$, resulting in a completely independent speckle pattern. 
The minimum value is $q=\tri\qf/2$ (see Section~\ref{secentropy}), in accordance with 
the uncertainty relation, and represents no visible change at all.
The exact relation between $q$ and the actual physical perturbation is hard to define. 
We will come back to this in Section~\ref{secexperiments}.

We use `hat' notation ($\;\hat{}\;$) for perturbed quantities, e.g. $\hat I(\vec x)$ 
is the intensity after perturbation.
Note that the statistical properties of $\hat\qa_{\vec a}$ are exactly the same as the 
properties of $\qa_{\vec a}$, i.e. uniformly distributed over the unit circle in the 
complex plane. Hence the statistics of $\hat I$ and $\hat G$ are the same as for the 
unperturbed quantities.
This is precisely what we want from our model, for there must be no `preference' in the 
formalism for either the perturbed or unperturbed speckle pattern.

Definition (\ref{defperturb}) allows us to study perturbations quantitatively, on a 
continuous scale as a function of~$q$.
Of special interest are the correlations between 
states before and after a perturbation.

We introduce the notation $\vwe{\cdot}$ for averaging with respect to the random 
variables $\{\qe_{\vec a}\}$.
We list several properties that will be useful in later sections.
The average effect on $\qa_{\vec a}$ is multiplication by the following factor:
\be
	\vwe{\exp i\qe_{\vec a}} = \intgr{-q}{q}{\qe}
	\frac{\exp i\qe}{2q}=
	\frac{\sin q}{q}.
\ee
From this the correlation between perturbed and unperturbed $\qa$ follows,
\be
	C(\hat\qa_{\vec a},\qa_{\vec b}):=
	\vwef{\hat\qa_{\vec a} \qa^*_{\vec b}}=
	\qd_{\vec a,\vec b}\frac{\sin q}{q}.
\ee
Here $\vwef{\cdot}$ indicates averaging first over $\qe$ and then over $\qf$.
Other useful identities are
\be
	\vwe{\exp i\nu\qe_{\vec a}}=\frac{\sin\nu q}{\nu q}
\ee
and
\be
	\vwe{\exp i(\qe_{\vec a}-\qe_{\vec b})}=
	\qd_{\vec a,\vec b}+(1-\qd_{\vec a,\vec b})
	\left(\frac{\sin q}{q}\right)^2.
\label{perturbexps}
\ee
For convenience later on we introduce the following shorthand notation,
\be
	Q:=\frac{\sin^2 q}{q^2}.
\ee
Note that $Q=1$ in the case of zero perturbation, and $Q=0$ when the perturbation has maximum strength ($q=\pi$).

\subsection{Mutual information in the source plane}
\label{secmutual}

The mutual information \cite{CT91} between two stochastic variables 
$X$ and $Y$ is denoted as ${\bf I}(X;Y)$.
It represents the amount of overlap in the information they carry, and it can be
computed as\footnote{
$H(X,Y)$ stands for the combined entropy of $X$ and $Y$.
The notation $H(Y|X)$ denotes the uncertainty in $Y$ given knowledge of $X$.
}
\be
	\bI(X;Y) = H(X)+H(Y)-H(X,Y) = H(Y) - H(Y|X).
\label{defmutual}
\ee
This quantity is of great importance in cryptography. Let $X$ and $Y$ be 
two noisy versions of the same secret, possessed by two parties respectively.
These parties can derive a common secret key from the variable that they hold. 
The theoretical maximum length of their common key is precisely given by the 
mutual information~$\bI(X;Y)$.

In our case, $X$ is the unperturbed speckle source $\qa$ and $Y$ is the perturbed 
source~$\hat\qa$. Their mutual information gives an absolute {\em physical} 
upper bound on the length of the key that can be derived from a speckle 
pattern {\em in a reproducible way}, for a given noise level~$q$.
The entropy $H[\qa]$ is given by~(\ref{entropy}).
The conditional entropy $H[\hat\qa |\qa]$ equals the uncertainty in the 
set~$\{\hat\qa_{\vec a}\}$ if $\{\qa_{\vec a}\}$ is known.
This is precisely the amount of information contained in the 
perturbation~$\{\qe_{\vec a}\}$. Thus we have
\be
	H[\hat\qa |\qa]= H[\qe] = \Nreg \log\frac{2q}{\tri\qf}.
\label{conditional}
\ee
Substitution of (\ref{entropy}) and (conditional) into (\ref{defmutual})
yields the result,
\be
	\bI(\hat\qa ; \qa)=\Nreg\log\frac{\pi}{q}.
\label{mutual}
\ee
Clearly, if $q$ has its maximum strength ($\pi$) then the mutual information is
zero. When $q$ has its minimum value, $\tri\qf/2$, then the mutual information
is equal to the full entropy (\ref{entropy}) of the source.

\begin{figure}[h!]
\begin{center}
\includegraphics[width=60mm]{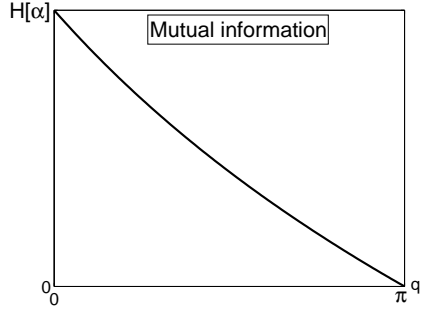}
\caption{\it Mutual information between unperturbed and perturbed source, 
as a function of the perturbation strength~$q$.}
\label{figmutualA}
\end{center}
\end{figure}

\subsection{Effect of perturbations on the intensity}
\label{secperturbI}

The effect of a perturbation on the intensity is computed as follows.  
First we average the perturbed intensity $\hat I(\vec x)$ over the perturbations 
$\qe_{\vec a}$. Using the representation (\ref{intensity}) and the identity (\ref{perturbexps}) we get
\be
	\vwe{\hat I(\vec x)}=
	QI(\vec x)
	+(1-Q)\Iav.
\label{vwIhat}
\ee
Eq.~(\ref{vwIhat}) shows how, on average, the perturbed value $\hat I(\vec x)$ gradually 
changes from $I(\vec x)$ to the average intensity $\Iav$ as a function of~$q$.

Next we compute the two-point correlation function between the perturbed and unperturbed speckle pattern.
Taking the expectation value $\vwf{\cdot}$ of (\ref{vwIhat}) multiplied with $I(\vec x')$ we obtain
\be
	C(\hat I(\vec x), I(\vec x')):=
	\frac{ \vwef{\hat I(\vec x) I(\vec x')} -\Iav^2}{\Iav^2}
	=
	Q C_I(\vec x, \vec x'),
\label{pertI2point}
\ee
where we have used the fact that the correlation with the constant number $\Iav$ in the 
last term of (\ref{vwIhat}) vanishes.
The result (\ref{pertI2point}) is intuitive as it states that the total correlation is 
the product of the ordinary two-point correlation $C_I$ (\ref{CIxxp}) and a perturbation effect.

The joint probability distribution $p(\hat I, I')$ is obtained by computing the moments \newline
$\vwef{ \hat I^n (I')^m }$ for all $n$ and~$m$.
This exercise is completely analogous to the derivation of~(\ref{momentInm}), but now with extra phase
factors $\exp i\qe_{\vec a}$.
Without showing the derivation, we mention that the averageing procedures
$\vwe{\cdot}$ and $\vwf{\cdot}$ in this computation commute
and give the result of the $\qf$-average,
\be
	\vwf{[\hat I(\vec x)]^n [I(\vec x')]^m}
	=\Iav^{n+m}n! m!\; {}_2 F_1\left(-n,-m;1; 
	\left| \frac{1}{\Nreg}\sum_{\vec a}e^{i\qe_{\vec a}} 
	e^{i\frac{\vec a\cdot(\vec x-\vec x')}{RM}}
	\right|^2\right).
\label{fullImoments}
\ee
Taking the $\qe$-average of (\ref{fullImoments}) is higly nontrivial in general.
However, an estimate with errors of order $1/\Nreg\ll 1$ is easily obtained by
replacing averages of powers by powers of averages,
\be
	\vwf{[\hat I(\vec x)]^n [I(\vec x')]^m}
	= \Iav^{n+m}n! m!\; {}_2 F_1\left(-n,-m;1;  \;
	Q C_I(\vec x,\vec x')\vphantom{\int}\right)
	\cdot\{ 1+{\cal O}(\frac{1}{\Nreg}) \}.
\ee
This gives, up to errors of order ${\cal O}(\Nreg^{-1})$, a joint probability distribution 
$p(\hat I(\vec x),  I(\vec x'))$ of the form (\ref{pIIp})
with $C_I$ replaced by~$Q C_I$.

\subsection{Effect of perturbations on the Gabor coefficients}
\label{secpertgabor}

The properties of the perturbed Gabor coefficients are readily computed.
We define the correlation between a perturbed coefficient $\hat G(s',\vec k',\vec x')$ and the 
unperturbed $G(s,\vec k,\vec x)$ as
\be
	C(G,\hat G'):=\frac{\vwef{G\hat G'}-
	\vwef{G}\vwef{\hat G'}}{\qs_G^2},
\label{defCGIMpert}
\ee
with $\qs_G$ as defined in~(\ref{sigG}).
We have again used shorthand notation $G$ for $G(s,\vec k,\vec x)$.
The averages in (\ref{defCGIMpert}) are easily evaluated.
First we note that $\vwf{G}=\vwf{\hat G}=0$.
Second, we factorize 
$\vwef{G\hat G'}=\vwf{G\vwe{\hat G'}}$.
It directly follws from the definition of $G$ (\ref{defG}) and the perturbation-
averaged intensity (\ref{vwIhat}) that
\be
	\vwe{\hat G'}=Q G'.
\label{vwhatG}
\ee
Thus we obtain
\be
	C(G,\hat G')= Q \cdot C_G(w,w',\vec k,\vec k',\vec
x,\vec x'),
\label{CGhatGp}
\ee
with $C_G$ as defined in (\ref{defCG}).
Hence, as in the case of intensity correlations, the correlation between 
the unperturbed $G$ at location $\vec x$ and the perturbed
$\hat G'$ at $\vec x'$ factorizes into a contribution from the perturbation and the 
ordinary correlation function~$C_G$.


\subsubsection{Mutual information between perturbed and unperturbed Gabor coefficients}
\label{secIGpert}

We estimate the mutual information
between an $n$-dimensional vector $G$ of unperturbed Gabor coefficients 
and the vector $\hat G$ of corresponding perturbed coefficients.
The vector consists of the same set as in Section~\ref{secdetectornoise}, i.e. a single width $w$,
two perpendicular wave vectors $\vec k_1$, $\vec k_2$ of equal length, and a grid
of points~$\vec x$.
We approximate the joint probability distribution of $G$ and $\hat G$ by a
Gaussian distribution.
This is motivated by the results of Section~\ref{sechighermoments}.

We make use of a well known result from information theory.
If $X$ is an $n$-component Gaussian-distributed 
vector with covariance matrix $\qS_X$, then the differential
entropy \cite{CT91} of $X$ is given by 
\be
	h(X)=\half\log_2 [(2\pi e)^n |\det \qS_X|].
\label{diffh}
\ee
Let $Y$ be a second vector of the same length, with covariance matrix $\qS_Y$,
and let $\qS_{XY}$ be the covariance of $X$ and~$Y$. 
Let $\qS_X$ and $\qS_{XY}$ commute.
Then the mutual information 
between $X$ and $Y$ follows from the definition (\ref{defmutual}) and (\ref{diffh}), 
\be
	\bI(X;Y)=-\half\log_2\left| 
	\det(1-\qS_Y^{-1}\qS_X^{-1}\qS_{XY}^2) 
	\vphantom{M^M}\right|.
\label{IXYgeneral}
\ee
Note that in the special case $Y=X+N$, where $N$ is Gaussian noise uncorrelated to
$X$, (\ref{IXYgeneral}) reduces to the form~(\ref{SNR}).

Now we consider the case where $X=G$ and $Y=\hat G+\qd G$, where $\qd G$ again is the Gaussian
detector noise discussed in Section~\ref{secdetectornoise}.
Note that $\hat G$ has the same $\qf$-ensemble statistics as~$G$.
Thus we have
\bea
	\qS_X\naar \qS_G, &
	\qS_Y\naar \qS_G+\qS_N, &
	\qS_{XY}\naar Q\qS_G.
\eea
Substitution into (\ref{IXYgeneral}) yields the mutual information
\be
	\bI(G;\hat G+\qd G)=\half\log_2\left|\det\left(
	1+\qS_N^{-1}\qS_G
	\right)\right|
	-\half\log_2\left|\det\left(
	1+[1-Q^2]\qS_N^{-1}\qS_G
	\right)\right|.
\label{IGhatGN}
\ee
This represents an absolute upper bound on the information that can be 
{\em reproducibly}
extracted from a
speckle pattern, given that there is noise $\qd I$ in the detector and 
perturbation noise $\{\qe_{\vec a}\}$ in the source. 
Clearly, (\ref{IGhatGN}) reduces to (\ref{SNR}) in the case $q\downarrow 0$, and
the mutual information goes to zero in the case $q\naar\pi$.

\begin{figure}[h!]
\begin{center}
\includegraphics[width=80mm]{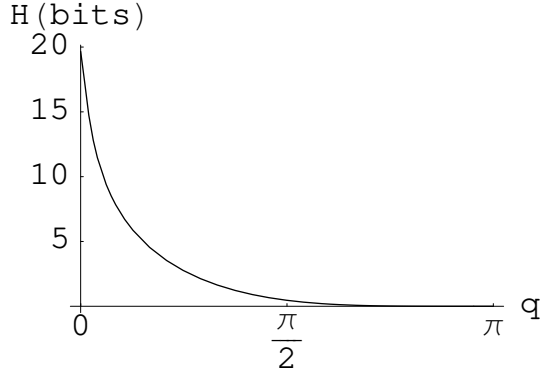}
\caption{\it Mutual information, per average speckle area, between a noiseless vector $G^{[j]}$ of Gabor coefficients and a noisy \& perturbed version
$\hat G^{[j]}+\qd G^{[j]}$, plotted as a function of the perturbation strength $q$. The parameters are: $L=800$ pixels, $M=3$ pixels, $\ell=5$ pixels, $t=1$ pixel\,$^2$, $\Iav/N_I=10$. The radius of an average speckle has been set, somewhat arbitrarily, to~$M$.}
\label{figmutualGhat}
\end{center}
\end{figure}

We estimate (\ref{IGhatGN}) using the approximation method given in~(\ref{psumapprox}).
This yields, for one direction of the $\vec k$-vector (and expressed in natural units instead of bits),
\bea
	\bI(G^{[j]};\hat G^{[j]}+\qd G^{[j]})  &\approx &
	\frac{\pi L^2}{2\ell^2}(1-\frac{\ell^2}{L^2})\ln\frac{1}{1-Q^2}
	\nn \\ &+&
	\frac{L^2}{2\pi c_2}\left[
	-\Dilog\frac{-e^{c_2 p^2}}{c_1}
	+\Dilog\frac{-e^{c_2 p^2}}{(1-Q^2)c_1}
	\right]_{p=\pi/L}^{\pi/\ell},
\label{dilogperturbed}
\eea
where the constants $c_1$ and $c_2$ are defined in~(\ref{defc1c2}).
The result (\ref{dilogperturbed}) is plotted in Fig.~\ref{figmutualGhat};
As in Fig.~\ref{figmutualG}, the factor $L^2$ was replaced by $\pi M^2$ to obtain a result per average speckle area.
The mutual information decreases sharply as a function of the perturbation strength~$q$.

\subsubsection{Bit error probability}
\label{secbitflip}

\begin{figure}[h!]
\begin{center}
\includegraphics[width=75mm]{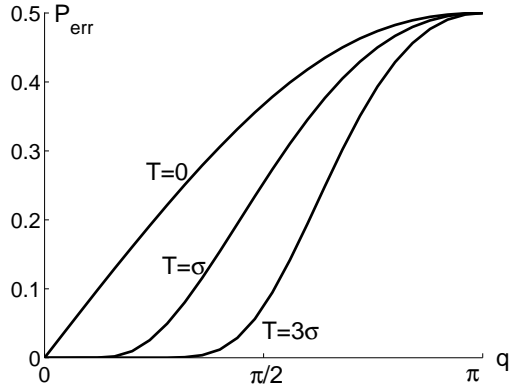}
\caption{\it Bit error probability $P_{\rm err}$ according to (\ref{errprobgen}) as a function of the perturbation strength~$q$ for various values of the threshold~$T$.}
\label{figerrorprob}
\end{center}
\end{figure}

We estimate
the probability of a bit error in a binarized Gabor coefficient
due to the random perturbation.
Here we will neglect the detector noise.
We use the shorthand notation $G=G(w,\vec k,\vec x)$ and 
$\hat G=\hat G(w,\vec k,\vec x)$.
As in the previous section, we approximate the joint probability distribution
$\qr$ of $G$ and $\hat G$ by a Gaussian,
making use of the correlation (\ref{CGhatGp}),
\bea
	\qr(G,\hat G)&=&\frac{1}{2\pi\sqrt{\det A}}\exp -\half
	(G,\hat G)A^{-1}\left(\matrix{G\cr \hat G}\right)
	\\
	A&=& \qs_G^2\left(\matrix{1&Q\cr 
	Q&1}\right).
	\nn
\eea
A bit error occurs when the signs of $G$ and $\hat G$ are not equal, while
 $|G|>T$ (see Section~\ref{secGabor}). The probability of this event is given by the following integral expression
\be
	P_{\rm err}=
	\prob{\hat G<0 | G>T}=\frac{\prob{\hat G<0, G>T}}{\prob{G>T}}
	=\frac{\intgr{-\infty}{0}{\hat G}\intgr{T}{\infty}{G}\qr}
	{\intgr{-\infty}{\infty}{\hat G}\intgr{T}{\infty}{G}\qr}.
\ee
Evaluation of several of the integrals gives
\bea
	P_{\rm err}&=&
	\frac{\intgr{T}{\infty}{G}\frac{1}{\qs_G\sqrt{2\pi}}
	e^{-\half G^2/\qs_G^2}\cdot
	\half\Erfc\frac{QG}{\sqrt{1-Q^2}\qs_G\sqrt{2}}}
	{\half\Erfc\frac{T}{\qs_G\sqrt{2}}},
\label{errprobgen}
\eea
where $\Erfc$ stands for the complementary error function.
Fig.~\ref{figerrorprob} shows the behaviour of $P_{\rm err}$ as a function of $q$ and~$T$.
Exact evaluation of the leftover integral in (\ref{errprobgen}) is difficult in general.
However, in some limiting cases analytic results can be obtained.
For instance, for $T=0$ the result is 
$P_{\rm err}=\pi^{-1}\arccos Q$.
Furthermore,
in the two limiting cases 
$\frac{Q}{\sqrt{1-Q^2}}\gg \frac{\qs_G\sqrt{2}}{T}$
and
$\frac{Q}{\sqrt{1-Q^2}}\ll \frac{\qs_G\sqrt{2}}{T}$
approximations can be obtained.
In the former case we apply a large-argument asymptotic expansion of the $\Erfc$ function; in the latter case a Taylor expansion.
For the first case, let us define the small parameter $\qe\ll 1$,
\be
	\qe:=\frac{\qs_G\sqrt{2}}{T}\frac{\sqrt{1-Q^2}}{Q}.
\ee
After some straightforward but tedious algebra we obtain
\be
	P_{\rm err}=\frac{e^{-1/\qe^2}e^{-\half T^2/\qs_G^2}}
	{\Erfc [T/(\qs_G\sqrt{2})]}
	\frac{T}{2\pi\qs_G\sqrt{2}} \left\{\qe^3
	-\qe^5(3/2 +\half T^2/\qs_G^2)
	+{\cal O}(\qe^7)\right\}.
\label{errorprobbig}
\ee
Expression (\ref{errorprobbig}) is useful in the weak perturbation limit $Q\naar 1$ and in the limit of large thresholds, $T/\qs_G \gg 1$.

For the second case we write $\qh:=1/\qe \ll 1$. A Taylor expansion of the $\Erfc$ function in (\ref{errprobgen}) yields, after some algebra,
\be
	P_{\rm err}=\frac{1}{2}\left[
	1-(\qh-\frac{\qh^3}{3\pi}[1+\frac{2\qs_G}{T}]) 
	\frac{\qs_G\sqrt{2}}{T}
	\frac{e^{-\half T^2/\qs_G^2}}{\half\Erfc[T/(\qs_G\sqrt{2})]}
	+{\cal O}(\qh^5)  \right].
\label{errorprobsmall}
\ee
Expression (\ref{errorprobsmall}) is useful in the strong perturbation limit $Q\naar 0$ and in the limit of small thresholds, $T/\qs_G\ll 1$.

\section{Comparison to experimental results}
\label{secexperiments}

In this section we briefly compare a number of theoretical results, obtained in the previous chapters, to actual experiments.
The experimental data were obtained with a very simple setup, consisting of a laser, a sample holder and a detector.
The laser has a wavelength of $\ql=$780nm. A parallel beam shines on the sample at an angle of 45$^\circ$. The spot is circular, with a diameter $D=$1mm. The sample is a piece of paper. The detector is a CCD camera, mounted at a normal angle to the sample. The distance between the sample and the camera is $z=$10cm. The camera has a pixel pitch of 6.25$\mu$m and takes 1024$\times$768 pixel images with 256 gray scales.
The typical speckle diameter at the location of the camera is of order
$\ql z/D=$78$\mu$m, corresponding to 12 pixels in the image.
The setup is not particularly well protected against background light.

\subsubsection*{Intensity distribution}
\begin{figure}[h!]
\begin{center}
\includegraphics[width=75mm]{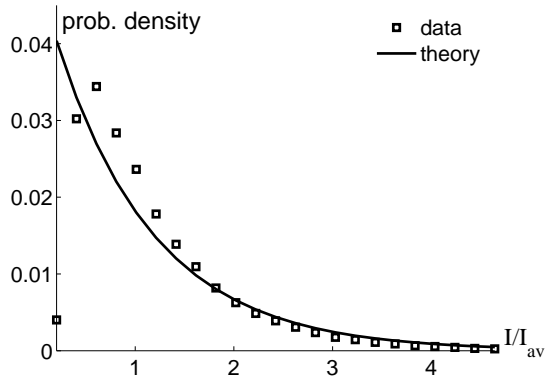}
\caption{\it Intensity histogram of a single speckle pattern. 
The histogram was made with a bin width of 5 gray values. 
The theoretical curve is the exponential distribution 
$(1/I_{\rm av})\exp(-I/I_{\rm av})$.}
\label{figIhist}
\end{center}
\end{figure}

In order to illustrate the quality of our data, we show
in Fig.~\ref{figIhist} a histogram of the gray levels present in a single typical CCD image. The lowest gray scale present in the image was normalised to zero.
The deviations from the theoretical curve (\ref{pI})
at low intensity show that there is a noticeable effect of the background light.

\subsubsection*{Statistical distribution of the Gabor coefficients}

Here we discuss the 
experimental verification of the theoretical results of Sections \ref{secGstat2ndorder} and~\ref{sechighermoments}. First we show that the theoretical prediction (\ref{sigG}) for $\qs_G$ is accurate. Then we show that the distribution function of the Gabor coefficients has a noticeable deviation from the Gaussian form, with fatter tails than a Gaussian.

The empirical $\qf$-ensemble probability distribution of $G(w,\vec k,\vec x_0)$
 should ideally be obtained as follows: Insert many  samples; for each sample, measure $G(w,\vec k,\vec x_0)$; finally make a histogram of all the Gabor coefficients; this yields the empirical distribution function for $G(w,\vec k,\vec x_0)$.

We used a less labour-intensive approach. We took a single sample; from the single CCD image we derived Gabor coefficients $G(w,\vec k,\vec x_0)$ for all $\vec x_0$; 
we made a histogram of the resulting set; we used this as the empirical distribution of $G(w,\vec k,\vec x_0)$.
This approach is motivated by (a) the fact that the $\qf$-ensemble probability distribution does not depend on $\vec x_0$
and (b) the ergodicity property of laser speckle, i.e. the property that 
the spatial intensity distribution asymptotically tends to 
the $\qf$-ensemble intensity distribution.

\begin{figure}[h!]
\begin{center}
\includegraphics[width=95mm]{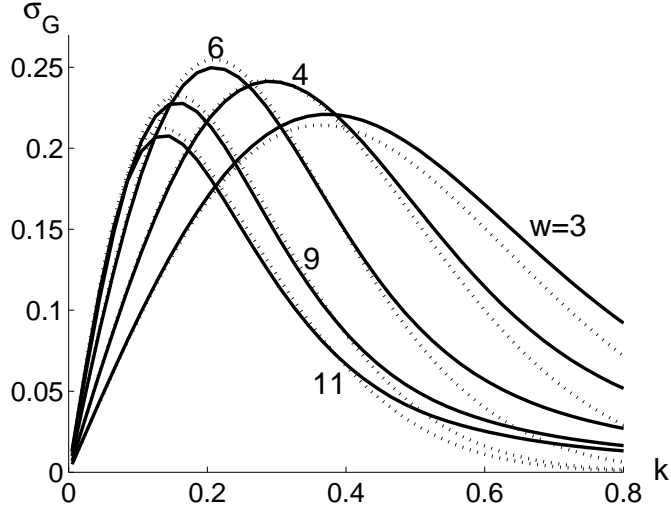}
\caption{\it The variance $\qs_G$ of the Gabor coefficients (normalized w.r.t. $\Iav$) as a function of the spatial frequency parameter $k$, for several values of the Gaussian width~$w$. (Here $w$ is measured in pixels, and $k$ in radians/pixel.)
The solid curves represent the experimental data. The dotted curves are the theoretical result~(\ref{sigG}), with the parameter choice $M=5$ pixels.}
\label{figsigmaGcurves}
\end{center}
\end{figure}

The result is shown in Fig.~\ref{figsigmaGcurves}, where $\qs_G$ is given as a function of $k=|\vec k|$ for a number of choices for~$w$.
The theoretical result (\ref{sigG}) is also plotted.
The correspondence of theory vs. experiment is very good, except at large~$k$. There we start to see the difference between the spatial continuum appraoch of the theory and the discrete pixellated nature of the CCD images. The theory uses spatial integration, while the data processing involves summation over pixels.
For fast oscillations, the integration in (\ref{defG}) 
averages out to zero more quickly than the summation.

\begin{figure}[h!]
\begin{center}
\includegraphics[width=70mm]{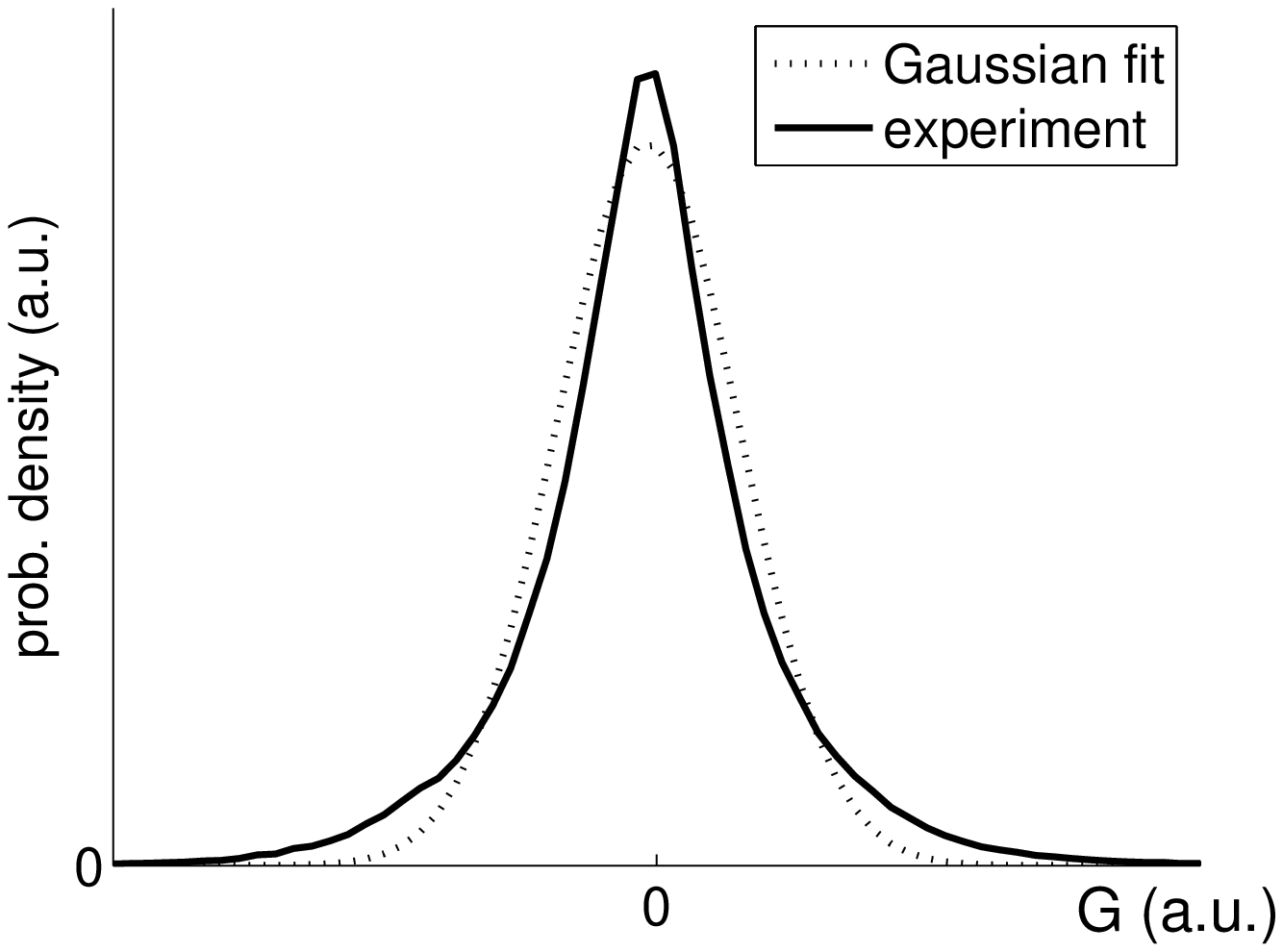}
\hskip5mm
\includegraphics[width=70mm]{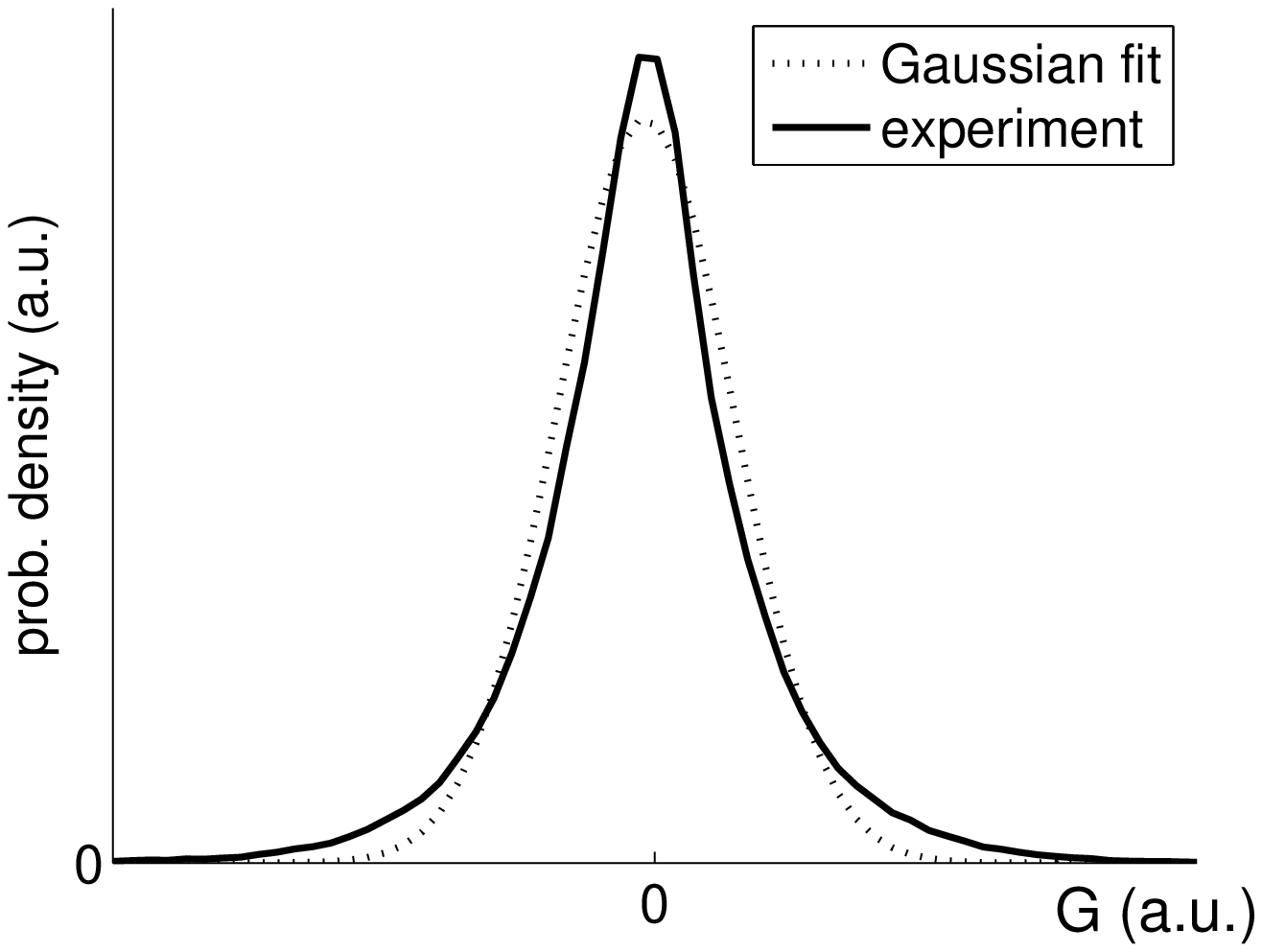}
\caption{\it Distribution of Gabor coefficients obtained from a single speckle pattern, by applying two perpendicular angle parameters. Left: $\qf=0$; Right: $\qf=90^\circ$.
The solid curve depicts a histogram of the Gabor coefficients. The dotted curve is the least-squares Gaussian fit to that histogram.}
\label{figGhist}
\end{center}
\end{figure}

Fig.~\ref{figGhist} shows the shape of the empirical distribution function of the Gabor coefficients. The curves were derived from a single image.
It can be seen that the distribution has fatter tails than a Gaussian, as was derived in Section~\ref{sechighermoments}.

\subsubsection*{Effect of perturbations}
\begin{figure}[h!]
\begin{center}
\includegraphics[width=95mm]{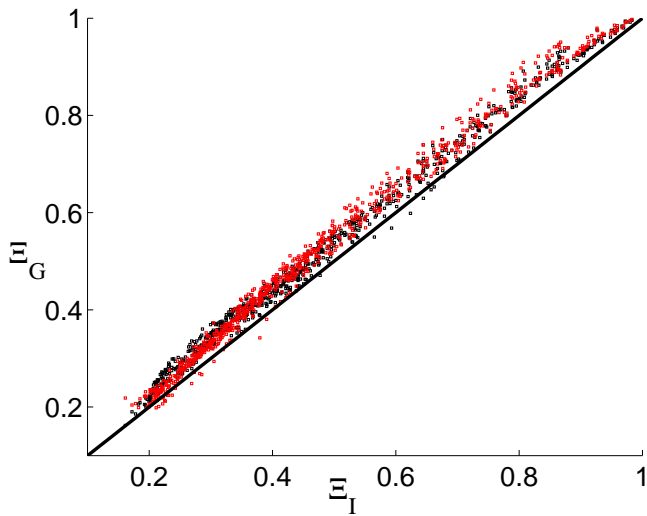}
\caption{\it Effect of a perturbation on the intensities and the Gabor coefficients. Horizontal axis: the intensity correlation function (\ref{XiI}). Vertical axis: The correlation function (\ref{XiG}) of the Gabor coefficients.
Data from two perpendicular $\vec k$ vectors are plotted together.
The solid line is the theoretical predicition for zero detector noise.
}
\label{figqcurves}
\end{center}
\end{figure}

We investigated perturbations as follows.
Our sample was a piece of paper whose surface structure changed over time.
We took 40 pictures at half hour intervals.
The changing surface structure can be regarded as a random perturbation as modelled in Section~\ref{secdefpert}.
Unfortunately, it is not possible to experimentally regulate the perturbation strength~$q$.
We therefore used the following approach to compare theory and experiment.
We looked at all pairs of CCD images; there are ${40 \choose 2}=780$ pairs.
For each pair $(A,B)$ we computed the empirical intensity correlation $\Xi_I$ and Gabor coefficient correlation $\Xi_G$,
\bea
\label{XiI}
	\Xi_I[A,B]&=&
	\frac{N_{\rm pix}^{-1}\sum_i I^A_i I^B_i
	-(N_{\rm pix}^{-1}\sum_i I^A_i)(N_{\rm pix}^{-1}\sum_i I^B_i)}
	{\sqrt{N_{\rm pix}^{-1}\sum_i [I^A_i]^2
	-[N_{\rm pix}^{-1}\sum_i I^A_i]^2}
	\sqrt{N_{\rm pix}^{-1}\sum_j[I^B_j]^2
	-[N_{\rm pix}^{-1}\sum_j I^B_j]^2}}
	\\
	\Xi_G[A,B]&=&
	\frac{N_{\rm pix}^{-1}\sum_i G^A_i G^B_i
	-(N_{\rm pix}^{-1}\sum_i G^A_i)(N_{\rm pix}^{-1}\sum_i G^B_i)}
	{\sqrt{N_{\rm pix}^{-1}\sum_i [G^A_i]^2
	-[N_{\rm pix}^{-1}\sum_i G^A_i]^2}
	\sqrt{N_{\rm pix}^{-1}\sum_j[G^B_j]^2
	-[N_{\rm pix}^{-1}\sum_j G^B_j]^2}}.
\label{XiG}
\eea
Here $N_{\rm pix}$ is the number of pixels in the image,
$\sum_i$ stands for summation over all pixels, $I^A_i$ denotes the intensity in image $A$ at location $\vec x_i$, and $G^A_i$ is shorthand notation for $G(w,\vec k,\vec x_i)$ for some fixed value of $w$ and $\vec k$.
The empirical correlation $\Xi_I$  should be equivalent to the theoretical correlation (\ref{pertI2point}) with $\vec x'=\vec x$;
Similarly, $\Xi_G$ should be equivalent to
(\ref{CGhatGp}) with the substitution $w'=w$, $\vec k'=\vec k$ and $\vec x'=\vec x$.
Hence, we expect $\Xi_I[A,B]=Q$ and $\Xi_G[A,B]=Q$.
In Fig.~\ref{figqcurves} we have plotted $\Xi_G$ vs. $\Xi_I$ for all image pairs. 
The data points are clearly bunched together on a narrow band slightly above the theoretically expected line~$\Xi_G=\Xi_I$.
We hypothesize that this small difference is due to detector noise, which we did not take into account here. The Gabor coefficients, resulting from a spatial sum, are less sensitive to detector noise than the intensity itself. Hence the correlation $\Xi_G$ of the Gabor coefficients is larger than the intensity correlation~$\Xi_I$.

\section{Summary}

Laser speckle has been proposed in the security literature as a source of high-entropy bit strings for various (cryptographic) purposes.
It is important to know how the physical properties of speckle affect the entropy of the extracted bit strings.
More in particular, we need to know the mutual information between two repeated key extractions when noise is taken into account.
Another important parameter is the bit error rate.

In this paper, we have developed a simple approach to address these issues.
We have studied the case of key extraction using Gabor coefficients. 
We used a simple model for speckle, generated by a large number of independent random phases in a source plane. 
We have modeled perturbations of the speckle pattern as small, uniformly distributed perturbations of the random phases.
Detector noise was modeled as being Gaussian, independent of the intensity and without correlations between the detector's pixels.

Our main results are
\begin{itemize}
\item
The Gabor coefficients have a distribution function that is close to Gaussian.
\item
We have derived an expression for the mutual information between an unperturbed and perturbed speckle source.
\item
We have obtained analytical expressions that give an upper bound on the mutual entropy of a set of Gabor coefficients 
(i) when there is detector noise but no perturbation
and
(ii) when there is detector noise as well as a perturbation.
\item
We have computed the bit error rate caused by perturbations of a speckle pattern.
\end{itemize}
Experimental data on the statistics of Gabor coefficients and on the correlation functions of Gabor coefficients and intensities are in accordance with theory.

The results of this paper, particularly the mutual information and error rate expressions, provide useful parameters for key extraction systems.

\subsubsection*{Acknowledgements}
We thank Sjoerd Stallinga, Geert-Jan Schrijen, Wil Ophey, Pim Tuyls, Frans Willems and Tanya Ignatenko for useful comments.

\bibliography{Gaborstats}

\appendix

\section{Fourth moment of the Gabor coefficients; case $w>M$}
\label{appG4}

In this appendix we calculate the non-Gaussian part of
$\vwf{G^4}$.
As discussed in section~\ref{secGstat}, (\ref{vwqanqan}) has 6 permutations with one cycle. 
These are equivalent because of the possibility of relabeling the dummy variables $\vec a_1,\cdots,\vec a_4$.
Introducing the notation
$S_{ij}=\sinh \frac{w^2}{MR} \vec k\cdot(\vec a_i -\vec a_j)$, we can write the non-Gaussian 
part of $\vwf{G^4}$ as
\bea
	&& 6\left[\frac{i\ql^2}{z^2}\right]^4
	e^{-2w^2 k^2} \sum_{\vec a_1,\cdots,\vec a_4}
	\exp\left[-\frac{w^2}{M^2 R^2} \sum_{j=1}^4 \vec a_j^2
	\right]
	\nn\\ &&
	\exp\left[ \frac{w^2}{M^2 R^2}(\vec a_1\cdot\vec a_2
	+\vec a_2\cdot\vec a_3+\vec a_3\cdot\vec a_4
	+\vec a_4\cdot\vec a_1)\right]
	S_{12}S_{23}S_{34}S_{41}.
\label{G4_1}
\eea
Note that the expressions in the exponents are invariant under relabeling of the summation variables.
This allows us to expand the product $S_{12}S_{23}S_{34}S_{41}$ into sixteen terms, 
which can then be grouped together into contributions of the same `type', i.e. equivalent under the summation.
This gives
\be
	S_{12}S_{23}S_{34}S_{41} \naar 
	\frac{1}{8}\left[1-2\cosh\frac{2w^2}{MR}\vec k\cdot
	(\vec a_1-\vec a_2)
	+\cosh \frac{2w^2}{MR}\vec k\cdot
	(\vec a_1-\vec a_2+\vec a_3-\vec a_4)\right].
\ee
Next we diagonalise the quadratic terms in the exponent of (\ref{G4_1}).
The exponent is of the form $\exp(-\qb^2 a \mu a^{\rm T})$, 
where $a$ denotes the four-component row vector $(\vec a_1,\cdots,\vec a_4)$, 
$\qb=w/(MR)$, and the matrix $\mu$ is given by
\be
	\mu=\left(\matrix{
	 \hphantom{-}1& -\half&  \hphantom{-}0& -\half\cr
	-\half&  \hphantom{-}1& -\half&  \hphantom{-}0\cr
	 \hphantom{-}0& -\half&  \hphantom{-}1& -\half\cr
	-\half&  \hphantom{-}0&  -\half& \hphantom{-}1
	}
	\right).
\ee
$\mu$ has one eigenvalue 0 with eigenvector $(1,1,1,1)$. The other three eigenvectors are
\bea
	\fr{1}{\sqrt{2}}(1,0,-1,0) &
	\frac{1}{\sqrt{2}}(0,1,0,-1) &
	\half(1,-1,1,-1)
\eea
with eigenvalues 1, 1 and 2 respectively.
We defining the new summation variables
$\vec v_0=\qb(\vec a_1+\vec a_2+\vec a_3+\vec a_4)$,
$\vec v_1=\qb(\vec a_1-\vec a_3)$,
$\vec v_2=\qb(\vec a_2-\vec a_4)$ and
$\vec v_3=\qb(\vec a_1-\vec a_2+\vec a_3-\vec a_4)$.
Next we approximate the summations by integrals: 
$\sum_{\vec a}\naar \ql^{-2}\int\!\rd^2 a$.
Taking the Jacobian into account, we write
$\int\!\rd^2 a_1\cdots\rd^2 a_4 = \qb^{-8} \int\!\rd^2 v_0\cdots\rd^2 v_3$.
Thus (\ref{G4_1}) can be approximated as
\bea
	&& \frac{3}{4}\qb^{-8}\ql^{-8}
	\left[\frac{i\ql^2}{z^2}\right]^4
	e^{-2w^2 k^2} 
	\int\!\rd^2 v_0\cdots\rd^2 v_3\;
	\exp\left[ -\half(\vec v_1^2+\vec v_2^2+\vec v_3^2) \right]
	\times \nn\\ &&
	[1+\cosh 2w\vec k\cdot\vec v_3
	-2\cosh w\vec k\cdot(\vec v_3+\vec v_1-\vec v_2)].
\label{uform}
\eea
Given the finite summation intervals of $\vec a_1\cdots \vec a_4$,
evaluation of the integrals in (\ref{uform}) does not yield esthetic results. 
However, if $w\gtrapprox 3M$ then a substantial part of the Gaussian distribution is covered by the integration, 
and considering the interval to be infinite is not a bad approximation. 
The case $w<M$ is discussed separately in Appendix~\ref{appG4_2}.
We only have to take into account the finiteness of the $v_0$-integral. 
The integrand in (\ref{uform}) does not depend on $\vec v_0$, and this leads to a factor 
$\pi(4\qb R)^2$, since $|\vec v_0|<4\qb R$.
The remaining integrals are readily evaluated. The final result is
\be
	\vwf{G^4}\approx 3\qs_G^4 +
	\frac{3}{2} \Iav^4 (\frac{M}{w})^6
	(1-2 e^{-w^2 k^2/2}+e^{-2w^2 k^2}).
\label{G4bigw}
\ee
From (\ref{sigG}) we see that $\qs_G^2$ is asymptotically proportional to $\Iav^2(M/w)^2$
for small $M/w$. Hence the result (\ref{G4bigw}) is of the form 
$3\qs_G^4(1+{\cal O}[M^2/w^2])$.

\section{Fourth moment of the Gabor coefficients; case $w\ll M$}
\label{appG4_2}

In this appendix we compute $\vwf{G^4}$ in the limit where the length scale $w$ 
of the Gabor transform is very small compared to the average speckle size.
In this limit, the intensity changes only slowly as a function of $\vec x$ 
within the Gaussian envelope. Around the point of interest $\vec x_0$ we can make a linear approximation
\be
	I(\vec x)\approx I(\vec x_0)+ \vec D\cdot(\vec x-\vec x_0),
\ee
with $\vec D=\nabla I(\vec x_0)$.
The Gabor transform (\ref{defG}) then reduces to
\be
	G(w,\vec k,\vec x_0) \approx w^2 \;\vec D\cdot\vec k
	\;e^{-w^2 k^2/2}.
\label{GIMsmall}
\ee
Differentiating (\ref{intensity}) and taking the inner product with $\vec k$, we obtain
\be
	\vec D\cdot \vec k =
	\frac{2\pi i\ql}{z^3}\sum_{\vec a,\vec b}\qa_{\vec a}
	\qa^*_{\vec b}\vec k\cdot(\vec a-\vec b)
	\exp\frac{i\pi}{\ql z}\left[\vec b^2-\vec a^2+2\vec x_0
	\cdot(\vec a-\vec b)\right].
\ee
We square (\ref{GIMsmall}), apply (\ref{vwqa4}) and replace the sums by integrals. In this way we obtain
\be
	\qs_G^2=\vw{G^2(w,\vec k,\vec x_0)}\approx
	4 w^4 k^2\Iav^2 M^{-2}e^{-w^2k^2}.
\ee
The fourth moment of (\ref{GIMsmall}) is obtained using (\ref{vwqanqan}) and again replacing summations by integrations,
\be
	\vw{G^4(w,\vec k,\vec x_0)}\approx 3\qs_G^4
	(1+\frac{1}{64}).
\ee

\end{document}